\pdfoutput=1
\documentclass[aps,prd,reprint,showpacs,preprintnumbers]{revtex4-1}

\usepackage{amsmath,amssymb}
\usepackage[bookmarks=true,bookmarksnumbered=true,
colorlinks=true,pdftitle={Paper},pdfauthor={Takuya Kanazawa, Tilo Wettig, Naoki Yamamoto},
pdfsubject={},pdfkeywords={}]{hyperref}

\renewcommand{\bar}[1]{\overline{#1}}
\renewcommand{\phi}{\varphi}
\renewcommand{\epsilon}{\varepsilon}

\newcommand{\ee}{{\,\textrm{e}}}
\newcommand{\muu}{\hat{\mu}}
\newcommand{\m}{\mathbf{m}}

\newcommand{\beq}{\begin{equation}}
\newcommand{\eeq}{\end{equation}}
\newcommand{\Sp}{\text{Sp}}
\newcommand{\SU}{\text{SU}}
\newcommand{\U}{\text{U}}
\DeclareMathOperator\tr{tr}
\DeclareMathOperator\Pf{Pf}
\DeclareMathOperator\re{Re}
\newcommand{\1}{\mathbf{1}}
\newcommand{\jfrac}[2]{\displaystyle\frac{#1}{#2}}
\newcommand\cc{\langle\bar\psi\psi\rangle}

\begin{document}
\preprint{TKYNT-09-25}
\title{Chiral random matrix theory for two-color QCD at high density}
\author{Takuya Kanazawa$^1$}
\author{Tilo Wettig$^2$}
\author{Naoki Yamamoto$^1$}
\affiliation{$^1$Department of Physics, The University of Tokyo, Tokyo
  113-0033, Japan} 
\affiliation{$^2$Department of Physics, University of Regensburg,
  93040 Regensburg, Germany} 
\date{April 21, 2010}

\begin{abstract}
  We identify a non-Hermitian chiral random matrix theory that
  corresponds to two-color QCD at high density.  We show that the
  partition function of the random matrix theory coincides with the
  partition function of the finite-volume effective theory at high
  density, and that the Leutwyler-Smilga-type spectral sum rules of
  the random matrix theory are identical to those derived from the effective
  theory.  The microscopic Dirac spectrum of the theory is governed by
  the BCS gap, rather than the conventional chiral condensate. We also
  show that with a different choice of a parameter the random matrix
  theory yields the effective partition function at low density.
\end{abstract}

\maketitle

Random matrix theory (RMT) has found numerous applications within and
outside of physics, e.g., in nuclear physics \cite{Wigner:1951},
mesoscopic condensed matter physics \cite{Guhr:1997ve}, quantum
chromodynamics (QCD) and QCD-like theories \cite{Verbaarschot:2000dy},
quantum gravity \cite{DiFrancesco:1993nw}, number theory
\cite{Conrey:2000}, econophysics \cite{Bouchaud:2009}, wireless
communications \cite{Tulino:2004}, and many more.  Most of these
applications are described by Hermitian random matrix models.  While
the first paper on non-Hermitian RMT by Ginibre appeared already in
1965 \cite{Ginibre:1965}, there was essentially no activity in this
area for a long time.  However, in the past 10 years or so numerous
new applications have been found in many fields of science that can be
described by non-Hermitian random matrix models \cite{[{For a
  comprehensive list of applications and references, see }]
  [{
  \href{http://www.comlab.ox.ac.uk/pseudospectra/applications.html}
   {http://www.comlab.ox.ac.uk/pseudospectra/applications.html}.
  }]
  pseudospectra}.  Prominent examples are QCD and QCD-like theories at
nonzero density which, depending on their anti-unitary symmetries, are
described by chiral versions of the Ginibre ensembles
\cite{Halasz:1997fc}.  So far, the applications of non-Hermitian RMT
in QCD have been restricted to moderate densities, roughly up to the
critical density for the chiral phase transition.  However, in recent
years there was quite some activity in the study of QCD and QCD-like
theories at very high densities, by means of effective theories and
renormalization group methods, and a number of interesting new
phenomena such as color superconductivity \cite{[{}][{ and references
    therein}]Alford2008} and color-flavor locking \cite{Alford:1998mk}
have been elucidated.  The present paper is the first application of
non-Hermitian RMT at high density and is expected to open up many more
applications.  We shall see that the situation is quite different
compared to the one at low density.  We will focus on the case of QCD
with two colors which can in principle be tested in lattice
simulations because the fermion sign problem is absent (for pairwise
degenerate quark masses).

Recently we have constructed the low-energy effective theory for
high-density QCD with two colors and an even number $N_f$ of fermions
in the fundamental representation \cite{Kanazawa:2009ks}.  At
sufficiently large quark chemical potential $\mu$ ($\mu\gg
\Lambda_{\SU(2)}$), chiral symmetry is broken spontaneously by a
diquark condensate as
\begin{align}
  \label{eq:symmetry}
  &\SU(N_f)_L \times \SU(N_f)_R \times \U(1)_B \times \U(1)_A \notag\\
  &\to \Sp(N_f)_L \times \Sp(N_f)_R\,,
\end{align}
and the degrees of freedom of the effective theory
are the Nambu-Goldstone (NG) bosons associated with (\ref{eq:symmetry}).

Considering the theory in a finite volume, we have
identified the corresponding microscopic regime (defined by the
requirement that the partition function is dominated by the
zero-momentum modes of the NG bosons) and derived a set of
Leutwyler-Smilga-type spectral sum rules \cite{Leutwyler:1992yt}
obeyed by the complex eigenvalues of the Dirac operator.  We
discovered that at high density the dimensionful quantity governing
the Dirac spectrum is not the usual chiral condensate but the BCS gap
$\Delta$ associated with the condensation of quark-quark pairs.  This
is also true for three-color QCD \cite{Yamamoto:2009ey}.  

The sum rules we obtained earlier plus a universality argument hinted
at the existence of a corresponding RMT.  Here, we
propose that the non-Hermitian chiral RMT describing two-color QCD at
large $\mu$ for an even number of fundamental fermions is given by
\begin{align}
  Z(\{m_f\})=\int dA\,dB~ & \ee^{-N \tr(AA^T+BB^T)}\notag\\
  & \times\prod_{f=1}^{N_f}
  \det\begin{pmatrix}m_f\1&A\\B&m_f\1\end{pmatrix},
  \label{eq:Z}
\end{align}
where $A$ and $B$ are real $N\times N$ matrices, the integration
measure is the flat Cartesian measure, the 
$m_f$ are dimensionless quantities corresponding to the quark masses, and $N_f$ is assumed to be even. 
At this stage physical scales have not been introduced yet.  (The
matching of random-matrix quantities to physical quantities is given 
in Eq.~\eqref{eq:identification} below.  There is no need to introduce
a Gaussian width parameter for $A$ and $B$, as this parameter can be absorbed in the $m_f$.)  Note that $A$ and $B$ are square
matrices.  In principle, the model \eqref{eq:Z} can be extended to
rectangular matrices $A$ and $B$ \cite{Akemann:2008mp}, see
Eq.~\eqref{eq:Zmu} below, giving rise to topological zero modes.
However, at high density the topological susceptibility is strongly
suppressed \cite{Schafer2002a, *Yamamoto:2008zw} so that topological
zero modes are irrelevant in the physical situation we are studying.

We stress that our model \eqref{eq:Z} differs from
earlier RMT approaches to QCD at $\mu\ne 0$ 
\cite{Vanderheyden2000,*Vanderheyden2000a,*Vanderheyden2001,Pepin2001,Klein2003,*Klein2005}.
In these approaches, the corresponding low-energy effective theory is formulated
in terms of NG bosons parametrizing the coset space $\SU(2N_f)/\Sp(2N_f)$ 
of the symmetry breaking pattern by a chiral condensate at $\mu=0$ \cite{Kogut2000}, i.e.,
$\mu$ is considered as a small perturbation.  On the other hand, the low-energy effective theory corresponding 
to \eqref{eq:Z} is formulated in terms of NG bosons parametrizing the coset space
$\SU(N_f) \times \SU(N_f)/[\Sp(N_f) \times \Sp(N_f)]$
(see \cite{Kanazawa:2009ks} and Eq.~\eqref{eq:Z_fin} below), i.e.,
it realizes the pattern of spontaneous chiral
symmetry breaking induced by BCS-type diquark condensation at high
density, with a vanishing chiral condensate.
To see this explicitly, we rewrite \eqref{eq:Z} in the chiral limit as
\begin{align}
  Z(\{0\}) & =\int \!dA~{\det}^{N_f/2} \begin{pmatrix}\mathbf{0}&A\\-A^T&\mathbf{0}\end{pmatrix}
  \ee^{-N\tr AA^T}
  \notag\\
  & \times \int \!dB~{\det}^{N_f/2}\begin{pmatrix}\mathbf{0}&B\\-B^T&\mathbf{0}\end{pmatrix}
  \ee^{-N\tr BB^T}\,.
  \label{eq:Z(0)}
\end{align}
Each of the two factors on the RHS of this equation corresponds to a
chiral orthogonal ensemble at $\mu=0$, for which the symmetry breaking
pattern is known to be $\SU(N_f) \to \Sp(N_f)$ \cite{Halasz1995a}.
Thus, we can immediately conclude that Eq.~\eqref{eq:Z(0)} exhibits
the symmetry breaking pattern $\SU(N_f) \times \SU(N_f) \to 
\Sp(N_f) \times \Sp(N_f)$, which agrees with Eq.~\eqref{eq:symmetry}.  
There are also NG bosons associated with the breaking of $\U(1)_B$ (baryon) and
$\U(1)_A$ (axial).  The former decouples completely and does not affect the quark
mass dependence of the partition function \cite{Kanazawa:2009ks}. 
The latter decouples in the chiral limit in which \eqref{eq:Z(0)} was considered. 
At nonzero quark masses, see \eqref{eq:Z}, we have a nontrivial $\U(1)_A$ integral, 
see Eq.~\eqref{eq:Z_fin} below.

The model \eqref{eq:Z} is not new.  It corresponds to class $2P$ in
Magnea's classification of non-Hermitian ensembles, see Table 2 of
\cite{Magnea:2007yk}.  
What is new is the realization that Eq.~\eqref{eq:Z} describes dense
two-color QCD.  Below we show explicitly that the effective theory
resulting from Eq.~\eqref{eq:Z} at large $N$ \emph{exactly reproduces}
the finite-volume partition function of dense two-color QCD in the
microscopic limit, obtained in \cite{Kanazawa:2009ks} based on
symmetry principles and weak-coupling calculations at large $\mu$.
This strongly suggests that Eq.~\eqref{eq:Z} is the correct RMT for
dense two-color QCD, paving the way towards a quantitative
understanding of the Dirac spectrum at large $\mu$.  A full proof of
the equivalence, which would also establish the equality of all
spectral correlation functions, requires the study of the partially
quenched theory, see \cite{Basile:2007ki}, which we leave to future
work.  However, we will also see that the Leutwyler-Smilga-type sum
rules derived from Eq.~\eqref{eq:Z} coincide with those obtained in
\cite{Kanazawa:2009ks}.  These sum rules are moments of the spectral
correlation functions and thus provide a nontrivial piece of evidence
for the equivalence.

Before going into details, let us list some basic properties of the
model \eqref{eq:Z}.  First, the matrix $\mathcal D\equiv
\begin{pmatrix}\mathbf{0}&A\\B&\mathbf{0}\end{pmatrix}$ is neither Hermitian nor
anti-Hermitian.  This implies that the eigenvalues of $\mathcal D$ are
distributed over the complex plane. Second, $\mathcal D$ preserves
chiral symmetry (i.e., it anticommutes with $\gamma_5$).
Third, it follows from chiral symmetry and from the fact that $A$ and
$B$ are real that the eigenvalues of $\mathcal D$ appear either in
pairs $(\lambda,-\lambda)$ with $\lambda\in \mathbb{R}\,\cup\,
i\mathbb{R}$ or in quartets $(\lambda,-\lambda,\lambda^*,-\lambda^*)$
with $\lambda\in\mathbb{C}$ \footnote{We expect the fraction of
  purely real or purely imaginary eigenvalues to go to zero with
  $1/\sqrt N$, and we have verified this expectation numerically for
  $N_f=0$.}.  All of the above features are in common with dense
two-color QCD \cite{Kanazawa:2009ks}.

Let us also elucidate the relation between our ensemble (\ref{eq:Z})
and related ensembles in the literature.  The first two-matrix model
for $\mu\ne0$, in which the matrix multiplying $\mu$ is not a unit
matrix but an independent random matrix, was introduced by Osborn
\cite{Osborn:2004rf}.  His original analysis was restricted to Dyson
index $\beta=2$, corresponding to three-color QCD.  Subsequently,
extensions to $\beta=4$ \cite{Akemann:2005fd} and $\beta=1$
\cite{Akemann:2008mp} were constructed.  Two-color QCD corresponds to
$\beta=1$, and the two-matrix model, which was studied mathematically
in \cite{Akemann:2009fc} for $N_f=0$, reads
\begin{align}
  Z_\nu(\muu,\{m_f\})& = 
  \int dC\,dD~\ee^{-2N\tr(CC^T+DD^T)}
  \notag \\
  &\quad \times \prod_{f=1}^{N_f} \det \begin{pmatrix}m_f\1 & C+ \muu
    D \\ -C^T+ \muu D^T & m_f\1 \end{pmatrix}. 
  \label{eq:Zmu}
\end{align}
Here, $C$ and $D$ are $N\times (N+\nu)$ real matrices and $\muu$ is a 
parameter corresponding to the physical chemical
potential $\mu$.  For $\nu=0$ and $\muu=1$, (\ref{eq:Zmu}) reduces to
(\ref{eq:Z}).

We shall now try to derive the effective theory corresponding to
\eqref{eq:Zmu} at large $N$ for $0\leq \muu \leq 1$, following a
similar analysis at $\muu=0$ \cite{Halasz1995a}.  The mass terms in
\eqref{eq:Zmu} are generalized to a generic mass term $\m^\dagger
P_R+\m P_L$ as in the QCD Lagrangian, where $\dim(\m)=N_f$
and $P_{R/L}=\frac12(1\pm\gamma_5)$.  We introduce Grassmann variables
to write the determinants as exponentials, integrate out the Gaussian
matrix elements, and then introduce auxiliary matrices for a
Hubbard-Stratonovich transformation. After integrating out the
Grassmann variables we obtain
\begin{align}
  & \!\!\! Z_\nu(\muu,\,\m) = \int dK\,dL\,dP~\ee^{-8 N \tr(KK^\dagger+LL^\dagger+2PP^\dagger)}
  \notag\\
  & \!\!\!\!\! \times \Pf^N
  \begin{pmatrix}
    \sqrt{1+\muu^2}L^\dagger & -\sqrt{1-\muu^2}P^T - \frac12\m^*
    \\
    \sqrt{1-\muu^2}P + \frac12\m^\dagger & \sqrt{1+\muu^2}K
  \end{pmatrix}
  \notag\\
  & \!\!\!\!\! \times \Pf^{N+\nu} \! 
  \begin{pmatrix}
    \sqrt{1+\muu^2}K^\dagger & \! \! -\sqrt{1-\muu^2}P^* - \frac12\m^T
    \\
    \sqrt{1-\muu^2}P^\dagger + \frac12\m & \! \! \sqrt{1+\muu^2}L
  \end{pmatrix},\!\!
  \label{eq:Z_mu}
\end{align}
where $K,\,L$ and $P$ are $N_f\times N_f$ complex-valued matrices, with 
$K$ and $L$ antisymmetric. 
$\Pf$ denotes the Pfaffian of the matrix. So far no approximation has been made.

The so-called strong non-Hermiticity limit is defined by $N\gg1$ with
$\muu$ fixed. We now focus on the case relevant for us, i.e., $\muu=1$ and $\nu=0$. 
In this case $P$ drops out and we find
\begin{align}
  Z_0(1,\m) 
   &= \int dK\,dL~\ee^{- N \tr(KK^\dagger+LL^\dagger)}
  \notag\\
  &\quad \times \Pf^N
  \begin{pmatrix}
    L^\dagger & -{\m^*}
    \\
    {\m^\dagger}& K
  \end{pmatrix}
  \Pf^{N}
  \begin{pmatrix}
    K^\dagger & -{\m^T}
    \\
    {\m}& L
  \end{pmatrix},
  \label{eq:Z0}
\end{align}
where the normalization has been changed slightly.

Below we assume that $N_f$ is even.  Since $K$ and $L$ are
antisymmetric they can be brought to the standard form $K = U\Lambda
U^T$ with $U \in \U(N_f)/[\Sp(2)]^{N_f/2}$ and $\Lambda$ a real
antisymmetric matrix with $\Lambda_{k,k+1}=-\Lambda_{k+1,k}\geq 0$ and
all other matrix elements zero (likewise for $L$). For $\|\m\| \ll 1$ 
the integration over $\Lambda$ can be estimated by a
saddle-point approximation at $\m=\mathbf{0}$, while the integration
over $U$ is soft and has to be done exactly at $\m\ne
\mathbf{0}$. Elementary calculation shows that the saddle point is
given by $\Lambda=I/\sqrt{2}$ with
$I=\begin{pmatrix}\mathbf{0}&\1\\-\1&\mathbf{0}\end{pmatrix}$.  Since
$U$ enters $Z_0$ only in the form $UIU^T$, the integration manifold
can be enlarged to $\U(N_f)$. Thus
\begin{align}
  & Z_0(1,\m)  \sim \!\!\! \int\limits_{\U(N_f)} \!\!\! dU\,dV~
  \Pf^N
  \begin{pmatrix}
    {\frac1{\sqrt{2}}(VIV^T)^\dagger} & -{\m^*}
    \\
    {\m^\dagger}& \frac1{{\sqrt{2}}}{UIU^T}
  \end{pmatrix}
  \notag\\
  & \!\! \qquad \qquad \qquad \qquad \quad \times
  \Pf^{N}
  \begin{pmatrix}
    {\frac1{\sqrt{2}}(UIU^T)^\dagger} & -{\m^T}
    \\
    {\m}& \frac1{\sqrt{2}}{VIV^T}
  \end{pmatrix}
  \notag
  \\
  & \sim \!\int\limits_{\U(N_f)}\!\!\! dU\,dV~\exp\big[-2N\re\tr (\m UIU^T \m^T V^*IV^\dagger)\big]\,,
\end{align}
where we expanded the exponent to the first nontrivial order in $\m$.
Redefining the variables as $U\equiv \tilde U \ee^{i(\theta+\phi)}$
and $V\equiv \tilde V\ee^{i(\theta-\phi)}$ with $\tilde U,\,\tilde
V\in \SU(N_f)$ and $0\le\phi\le\pi$, and setting $A\equiv \ee^{2i\phi}$, we obtain 
\begin{align}
  &Z_0(1,\m) \sim \int\limits_{\U(1)}\! dA \!\!
  \int\limits_{\SU(N_f)} \!\!\! d\tilde{U}\,d\tilde{V} \notag\\
  & \quad\times \exp\big[-2N\re\tr (A^2 \m \tilde{U}I\tilde{U}^T
  \m^T \tilde{V}^*I\tilde{V}^\dagger)\big]\,, 
  \label{eq:Z_fin}
\end{align}
which exactly coincides with the
finite-volume partition
function at large $\mu$ \cite[Eq.~(4.2), Eq.~(4.24)]{Kanazawa:2009ks}
if we identify
\begin{equation}
  N\,\m^2 \Longleftrightarrow
  \jfrac{3}{4\pi^2}V_4\Delta^2M^2\,,
  \label{eq:identification}
\end{equation}
where $M$ is the dimensionful physical mass matrix and $V_4$ is the
space-time volume.  Therefore, our model (\ref{eq:Z}) depends 
on the chemical potential $\mu$ only implicitly through
$\Delta$. (Note that the gap $\Delta$ depends on $\mu$ via the relation
    $\Delta \sim \mu \ee^{-1/g}$ \cite{Son1999a} and that 
$\Delta$ is related to the diquark condensate via the relation 
$\langle qq\rangle \sim \mu^2\Delta/g$ \cite{Schafer2002,Kanazawa:2009ks}.) 
The absence of terms linear in $\m$ in the exponent of
(\ref{eq:Z_fin}) is a consequence of the $\mathbb{Z}(2)_L \times
\mathbb{Z}(2)_R$ symmetry (flipping left-handed and right-handed
quarks independently, i.e., $q_L \rightarrow \pm q_L$ and $q_R
\rightarrow \pm q_R$) of the diquark pairing $\langle q_L q_L \rangle$
and $\langle q_R q_R \rangle$, which distinguishes the model
\eqref{eq:Z} from conventional chiral RMTs.
The physical limits in which the model \eqref{eq:Z} describes
two-color QCD are given by the inequalities defining the microscopic
domain (or lowest order of the $\varepsilon$-regime) at high density
\cite{Kanazawa:2009ks},
\begin{align}
  \frac1\Delta\ll L\ll\frac1{m_{\Pi,\bar\Pi,\eta'}}\,,
\end{align}
where $L$ is the linear extent of the Euclidean box and
$m_{\Pi,\bar\Pi,\eta'}$ is the mass of the NG bosons at high density.

In addition to the mass term, we could also add a diquark source term $\propto jqq$ 
to the model \eqref{eq:Z}. It can be shown that the partition function at $N\gg 1$ 
then contains a term linear in $j$, resulting in a nonvanishing diquark condensate 
in this model. The addition of the diquark source term makes the model somewhat more complicated and will be addressed 
in future work \cite{[{}][{ in preparation.}]In_preparation}.

Since in the large-$N$ limit the partition function \eqref{eq:Z} has
the same mass dependence as the partition function \eqref{eq:Z_fin} of
the effective theory, it follows immediately from the calculations
performed in \cite{Kanazawa:2009ks} that the Leutwyler-Smilga-type sum
rules for inverse Dirac eigenvalues of the model \eqref{eq:Z} agree
with those obtained in \cite{Kanazawa:2009ks}, as stated earlier.

For $N_f=2$ we can also calculate the partition function at finite
$N$. In \cite{Akemann:2008mp} the correlation
function of two characteristic polynomials was studied in the quenched version of
the model (\ref{eq:Zmu}), which corresponds exactly to $Z_\nu$ for $N_f=2$
in (\ref{eq:Zmu}). The result for $\muu=1$ (after a 
change of notation and correcting typos) reads
\cite[Eq.~(34)]{Akemann:2008mp}
\begin{align}
  Z_\nu(1,\{m_1,m_2\})
  \propto & \ \sum_{\ell=0}^{N} \frac{1}{\ell!(\ell+\nu)!}
  (2Nm_1m_2)^{2\ell + \nu}  \notag\\
  \to & \ I_\nu(4Nm_1m_2) \text{ as } N\to \infty\,.
\end{align}
The final expression $I_0(4Nm_1m_2)$ can be obtained from
(\ref{eq:Z_fin}) as well, 
thus verifying our saddle-point result.

We now comment on the case of odd $N_f$. With $\muu=1$, assuming
$\m=m\1$ for simplicity, and allowing for $\nu\ne0$ again, we find for
the Pfaffians in the integrand of \eqref{eq:Z0}
\begin{align}
  \label{eq:antisym}
  & \Pf^N
  \begin{pmatrix}
    L^\dagger & -{\m^*} \\
    {\m^\dagger}& K
  \end{pmatrix}
  \Pf^{N+\nu}
  \begin{pmatrix}
    K^\dagger & -{\m^T} \\
    {\m}& L
  \end{pmatrix}\\
  & \propto {\det}^{N/2}({m^*}^2\1+L^\dagger K) \cdot
  {\det}^{(N+\nu)/2}(m^2\1+LK^\dagger)\,.\notag 
\end{align}
Since $K$ (or $L$) is an antisymmetric matrix of odd size, at least
one of its eigenvalues must be zero, and evidently the same holds for
$L^\dagger K$ and $LK^\dagger$. Thus 
$Z_\nu(1,\m) \propto |m|^{2N} m^\nu$, and
hence the model with odd $N_f$ seems to have an ill-defined
chiral limit, with a divergent chiral condensate. 
Further analysis is left to future work.

While so far we have mainly been concerned with the $\muu=1$ case,
there exists another region of $\muu$ which is nontrivial, yet
analytically tractable: the weak non-Hermiticity limit ($N\gg 1$ with
$N\muu^2$ fixed) \cite{Fyodorov:1996sx}, to which we now turn for the
sake of completeness. First, define
\begin{equation}
  \mathfrak D\equiv 
  \begin{pmatrix} 
    K^\dagger & -P^* \\ 
    P^\dagger & L 
  \end{pmatrix} 
  \quad\text{and}\quad 
  \mathfrak M\equiv 
  \begin{pmatrix}
    \mathbf{0}&-\m^T\\
    \m&\mathbf{0}
  \end{pmatrix}\,.
\end{equation}
In the following discussion $N_f$ is not restricted to be even.  Since
\emph{both} $\muu^2$ and $\m$ are treated as infinitesimal parameters,
the saddle point is estimated at $\muu=0$ and $\m=\mathbf{0}$, i.e., $\mathfrak
D=UIU^T/4$ with $U\in\U(2N_f)$.  Taylor expansion in $\muu^2$ yields
\begin{align}
  & \Pf
  \begin{pmatrix}
    \sqrt{1+\muu^2}K^\dagger & -\sqrt{1-\muu^2}P^* - \frac12\m^T
    \\
    \sqrt{1-\muu^2}P^\dagger + \frac12\m & \sqrt{1+\muu^2}L
  \end{pmatrix}
  \\
  & \sim \det U \exp\left[4\muu^2 \tr (\mathfrak D \mathfrak B^T \mathfrak D^\dagger \mathfrak B)
  + 4\tr(\mathfrak D^\dagger \mathfrak M) \right]\,,
\notag
\end{align}
where we used $\mathfrak D^\dagger \mathfrak D=\1/16$ and
introduced the $2N_f\times 2N_f$ baryon charge matrix $\displaystyle
\mathfrak B\equiv \begin{pmatrix}\1&\mathbf{0}\\\mathbf{0}&-\1\end{pmatrix}$.
Isolating the $\U(1)$ part by $U \to U\ee^{i\theta/2N_f}$ we find
\begin{align}
  Z_\nu(\muu,\m)&\sim
  \int d\theta \, \ee^{i\nu\theta} \hspace{-9pt}
  \int\limits_{\SU(2N_f)}\hspace{-9pt} dU~ \notag \\
  & \times \exp\Big[
  \frac{1}{2} N \muu^2 \tr \Big(UIU^T \mathfrak B^T (UIU^T)^\dagger \mathfrak B\Big)
  \notag
  \\
  & \qquad \quad + 2N \re \tr (\ee^{i\theta/N_f} UIU^T \mathfrak
  M^\dagger) \Big]\,.
  \label{eq:Z_2-color_small_mu}
\end{align}
The exponent in this partition function exactly reproduces the static
part of the chiral Lagrangian \cite[Eq.~(42)]{Kogut2000} obtained from
symmetry principles, with the identification $\frac{1}{2} N\muu^2
\equiv V_4 F^2\mu^2$ and $2N m \equiv V_4F^2m_\pi^2=V_4|\cc|M/2N_f$, where $F$ is
the usual low-energy constant, $m_\pi$ is the pion mass, and in the last
equality the Gell-Mann--Oakes--Renner relation was used. (A similar analysis
is given in \cite[Eq.~(4.15) of 2nd reference]{Klein2005} based on the
one-matrix formulation.)  
It is intriguing that a \emph{single} chiral
RMT, Eq.~(\ref{eq:Zmu}), can describe two extreme cases, $\mu\ll
\Lambda_{\SU(2)}$ and $\mu\gg \Lambda_{\SU(2)}$, that have totally
distinct patterns of spontaneous symmetry breaking, by two different
choices of the parameter ($\muu \sim O(1/\sqrt{N})$ and $\muu=1$,
respectively) and two different mappings of the random-matrix quark masses to the
physical quark masses (rescaling by $|\cc|$ and $\Delta$, repectively).

We expect that this paper will stimulate work in several directions.
(i) While we do not doubt that \eqref{eq:Z} yields the correct
microscopic spectral correlation functions for two-color QCD at high
density, for completeness it would be useful to prove the equivalence
rigorously using the partially quenched theory.  (ii) We are currently
investigating the calculation of the microscopic spectral correlation
functions using the methods developed in \cite{Akemann:2009fc}.  (iii) 
It would be stimulating to extend our model to two-color QCD at 
intermediate densities, the latter being relevant to the phase
structure of two-color QCD \cite{Splittorff2001,Klein2005}. The issue of 
continuity between the hadronic phase and the BCS superfluid phase is of 
particular interest, in view of the formal similarity of the partition 
functions at small and large $\mu$ for degenerate masses \cite{Kanazawa:2009ks}. 
(iv) Since two-color QCD with even $N_f$ and pairwise
degenerate quark masses is
free from the sign problem even at $\mu\ne 0$, it is in principle
possible to test the predictions of our model by lattice simulations. 
(v) Finally, the
most important extension from a phenomenological viewpoint is to the
color-superconducting phase of three-color QCD \cite{Yamamoto:2009ey}.
Work in most of these directions is in progress.

\begin{acknowledgments}
  We thank G.~Akemann for useful correspondence and for providing us
  with Ref.~\cite{Akemann:2009fc} prior to publication, and
  J.~J.~M.~Verbaarschot  for helpful comments on
  the manuscript.  TK and NY are
  supported by the Japan Society for the Promotion of Science for
  Young Scientists.  TW acknowledges support by DFG.
\end{acknowledgments}

\bibliography{paper}

\end{document}